\documentclass[aps,prd,reprint,superscriptaddress,nofootinbib,floatfix,showkeys]{revtex4-2}

\usepackage{graphicx}
\usepackage{amsmath}
\usepackage{amsfonts}
\usepackage{amssymb}
\usepackage{xcolor}
\usepackage{times,txfonts}
\usepackage{mathtools}
\usepackage[hidelinks,pdfusetitle,hypertexnames=false]{hyperref}
\usepackage{ulem}

\usepackage{physics}
\usepackage{enumitem}
\usepackage{hyperref}
\hypersetup{
  colorlinks=true,
  linkcolor=blue,
  citecolor=cyan,
}
\usepackage{orcidlink}

\newcommand{\e}{\mathrm{e}}
\newcommand{\ii}{\mathrm{i}}
\newcommand{\calF}{\mathcal{F}}
\newcommand{\calM}{\mathcal{M}}
\newcommand{\Rs}{R_{\star}}
\newcommand{\fs}{f_{\star}}
\newcommand{\rstar}{r_{*}}
\newcommand{\pstar}{\partial_{r_*}}
\newcommand{\Wight}{W}
\newcommand{\Order}{\mathcal{O}}

\definecolor{coauthorgreen}{RGB}{0,128,0}

\begin{document}


\title{Boundary-induced corrections to the response of an Unruh–DeWitt detector outside a compact star with Robin surface conditions}


\author{Nodirbek Anarboev}
\email{nodirbekanarboyev@gmail.com}
\affiliation{Institute of Theoretical Physics, National University of Uzbekistan, Tashkent 100174, Uzbekistan}

\author{Javlonbek Khasanov\orcidlink{0000-0001-9293-1838}}
\email{khasanovjavlonbek20@gmail.com}
\affiliation{Institute of Theoretical Physics, National University of Uzbekistan, Tashkent 100174, Uzbekistan}

\author{Javlon~Rayimbaev~\orcidlink{0000-0001-9293-1838}}
\email{javlon@astrin.uz}
\affiliation{University of Tashkent for Applied Sciences, Gavhar Str. 1, Tashkent 700127, Uzbekistan}
\affiliation{Kimyo International University in Tashkent, Shota Rustaveli Street 156, Tashkent 100121, Uzbekistan}

\author{Bobomurat Ahmedov\orcidlink{0000-0002-1232-610X}} 
\email{ahmedov@astrin.uz}
\affiliation{School of Physics, Harbin Institute of Technology, Harbin 150001, People’s Republic of China}
\affiliation{Institute for Advanced Studies, New Uzbekistan University, Movarounnahr str. 1, Tashkent 100000, Uzbekistan}

\begin{abstract}
We formulate the finite-time response problem for a radially accelerated Unruh--DeWitt detector coupled to a real massless scalar field outside a static, spherically symmetric, horizonless compact star. The stellar exterior is described by the Schwarzschild geometry, while unresolved field--surface interactions are represented by a local Robin condition at the stellar radius.  After reducing the Klein--Gordon equation to a half-line Schr\"odinger problem, we identify a sufficiently stable branch of the Robin family from positivity of the radial operator, construct delta-normalized continuum modes, and define the static ground state with respect to Schwarzschild time.  We derive a mode-amplitude representation of the smoothly switched detector response that avoids a direct numerical treatment of the coincidence singularity of the Wightman distribution.  We obtain a radially aligned trajectory of constant signed proper acceleration in terms of the rapidity measured by static observers.  The resulting formulation separates the effects of compactness, surface condition, closest approach, detector gap, and interaction duration.  We also state the flat-space, Dirichlet, static-detector, low-frequency, large-distance, and long-interaction limits required to validate a numerical implementation. The framework provides a controlled extension of detector studies with static reflective compact objects to finite-time moving probes.\\[4 pt]
\end{abstract}

\keywords{Unruh--DeWitt detector, Unruh effect, compact star, Robin boundary condition, Schwarzschild spacetime, finite-time detector response, quantum field theory in curved spacetime, finite-time response, Schwarzschild exterior}

\maketitle

\section{Introduction}
\label{sec:introduction}

The Unruh effect gives an operational meaning to the observer dependence of particle content: a uniformly accelerated detector in the Minkowski vacuum has a thermal response at the temperature $T_{\mathrm U}=a/(2\pi)$ in natural units~\cite{Unruh1976,BirrellDavies1982}.  The Unruh--DeWitt (UDW) detector idealizes the probe as a localized quantum system coupled to a scalar field~\cite{DeWitt1979}.  Beyond stationary settings, the response depends not only on the trajectory and the quantum state but also on how the interaction is switched on and off.  Smooth switching is essential for a regulator-independent finite-time response in a Hadamard state~\cite{Satz2007,LoukoSatz2008}.

Boundaries and defects alter the mode spectrum and, hence, the correlation function a detector samples.  In static spacetimes with a timelike boundary, physically admissible dynamics are tied to positive self-adjoint extensions of the spatial wave operator~\cite{Wald1980,IshibashiWald2003}.  Robin conditions provide a natural one-parameter realization of such extensions, but positivity and infrared regularity are not automatic for every parameter choice~\cite{CamposDappiaggiSinibaldi2024}.  These issues are especially relevant when a boundary-induced contribution is separated from a reference two-point function.

Researchers have investigated UDW detectors in Schwarzschild spacetime numerically for static and circular-geodesic trajectories in the Boulware, Hartle--Hawking, and Unruh states~\cite{HodgkinsonLoukoOttewill2014}.  Researchers have also developed dimensionally reduced detector models for general static spherically symmetric geometries~\cite{TjoaMann2022}.  For horizonless exotic compact objects (ECOs), Holdom, Mann, and Zhang studied the stationary response of a static detector in the presence of a general reflective surface and identified resonance structures associated with the cavity between the surface and the curvature potential~\cite{HoldomMannZhang2021}.  Their construction focused on an ECO surface close to the would-be horizon and on a time-independent transition rate.

More recently, Botta, Pitelli, and Mosna considered a uniformly accelerated detector coupled to a scalar field on punctured Minkowski spacetime with a Robin family at the removed origin~\cite{BottaPitelliMosna2026}.  The puncture breaks boost invariance, making the boundary-induced pullback nonstationary.  For a finite Robin parameter, the corresponding finite-time correction remains $\Order(1)$ in their long-interaction limit, whereas the formal Neumann endpoint has an infrared singularity.  A finite stellar surface in a curved exterior, however, is not obtained by simply importing the puncture modes: all angular sectors can scatter from the surface, the Schwarzschild curvature produces an effective potential, and the accelerated trajectory is no longer an orbit of a global boost symmetry.

In this work, we therefore formulate a different problem: a smoothly switched, radially accelerated detector outside a nonrotating compact star whose unresolved surface interaction is represented by a local Robin condition.  The principal distinctions from Ref.~\cite{HoldomMannZhang2021} are the moving detector, finite interaction time, a local self-adjoint Robin condition rather than a prescribed constant reflectivity, and stellar radii not restricted to the near-horizon ECO regime.  Our main observable is the Robin deformation relative to a Dirichlet surface on the same exterior domain,
\begin{equation}
  \Delta\calF_{\Gamma D}(\Omega)
  :=\calF_{\Gamma}(\Omega)-\calF_D(\Omega),
  \label{eq:response_difference_intro}
\end{equation}
where $\Gamma$ denotes the reduced Robin coefficient and $\Omega$ is the detector energy gap.  This subtraction isolates the change in surface conditions without conflating it with a change in the spacetime domain.

The paper is organized as follows.  Section~\ref{sec:geometry} defines the exterior geometry and obtains the radial Klein--Gordon equation.  Section~\ref{sec:robin} relates the covariant surface condition to a Robin condition for the reduced radial field and identifies a sufficiently stable branch.  Section~\ref{sec:quantization} constructs normalized modes and the static ground-state Wightman function.  Section~\ref{sec:trajectory} derives a radial trajectory of constant proper acceleration.  Sec.~\ref{sec:detector} derives the finite-time detector response and a numerically stable mode-amplitude representation.  Sec.~\ref {sec:limits} and~\ref{sec:numerics} give analytic limits and the numerical program, followed by the observables evaluated in Sec.~\ref{sec:results}.

Throughout we use units $G=c=\hbar=1$ and metric signature $(-,+,+,+)$.

\section{Exterior geometry and scalar-field dynamics}
\label{sec:geometry}

We consider the exterior domain
\begin{equation}
  \calM_{\mathrm{ext}}=\{(t,r,\theta,\varphi):r\geq\Rs\},
  \qquad \Rs>2M,
  \label{eq:domain}
\end{equation}
of a static, nonrotating, horizonless compact star of gravitational mass $M$ and areal radius $\Rs$.  By spherical symmetry and vacuum outside the material body, the exterior line element is
\begin{equation}
  \dd s^2=-f(r)\,\dd t^2+\frac{\dd r^2}{f(r)}
  +r^2\left(\dd\theta^2+\sin^2\theta\,\dd\varphi^2\right),
  \label{eq:schwarzschild_metric}
\end{equation}
where  $f(r)=1-\cfrac{2M}{r}$. We treat the surface $r=\Rs$ as an effective timelike boundary.  This model is appropriate when, over the frequency range of interest, the probe field's microscopic penetration and interaction with the stellar material can be replaced by a conservative local response.  A complex or explicitly frequency-dependent surface coefficient would describe absorption or memory and lies outside the present closed-system model.

The bulk action for a real scalar field is
\begin{equation}
  S_{\mathrm{bulk}}
  =-\frac12\int_{\calM_{\mathrm{ext}}}\dd^4x\sqrt{-g}
  \left(g^{\mu\nu}\nabla_\mu\Phi\nabla_\nu\Phi
  +\mu_s^2\Phi^2+\xi\mathcal{R}\Phi^2\right).
  \label{eq:bulk_action}
\end{equation}
The Schwarzschild exterior is Ricci flat, $\mathcal{R}=0$.  We specialize to a massless, minimally coupled field,
\begin{equation}
  \mu_s=0,
  \qquad \xi=0,
  \qquad \Box\Phi=0.
  \label{eq:kg}
\end{equation}

Introduce the tortoise coordinate
\begin{equation}
  \frac{\dd\rstar}{\dd r}=\frac{1}{f(r)},
  \qquad
  \rstar=r+2M\ln\left(\frac{r}{2M}-1\right),
  \label{eq:tortoise}
\end{equation}
where an irrelevant additive constant has been fixed by convention.  Separating variables as
\begin{equation}
  u_{\omega\ell m}(x)
  =\frac{1}{\sqrt{2\omega}}\,
  \e^{-\ii\omega t}Y_{\ell m}(\theta,\varphi)
  \frac{\psi_{\omega\ell}(r)}{r},
  \qquad \omega>0,
  \label{eq:separated_mode}
\end{equation}
reduces Eq.~\eqref{eq:kg} to
\begin{equation}
  \left[-\frac{\dd^2}{\dd\rstar^2}+V_\ell(r)\right]
  \psi_{\omega\ell}(r)=\omega^2\psi_{\omega\ell}(r),
  \label{eq:radial_schrodinger}
\end{equation}
with
\begin{equation}
  V_\ell(r)=f(r)
  \left[\frac{\ell(\ell+1)}{r^2}+\frac{2M}{r^3}\right].
  \label{eq:potential}
\end{equation}
The potential is nonnegative for $r>2M$ and vanishes at spatial infinity.  In the exterior-star problem Eq.~\eqref{eq:radial_schrodinger} is a half-line scattering problem on $\rstar\in[\rstar(\Rs),\infty)$.

For later numerical work, it is useful to write the same equation in the areal coordinate:
\begin{equation}
  f^2\frac{\dd^2\psi_{\omega\ell}}{\dd r^2}
  +ff'\frac{\dd\psi_{\omega\ell}}{\dd r}
  +\left(\omega^2-V_\ell\right)\psi_{\omega\ell}=0.
  \label{eq:radial_r}
\end{equation}

The static sector supplies a useful analytic check.  Writing the radial part
of the unreduced field as $R_\ell=\psi_{0\ell}/r$, Eq.~\eqref{eq:kg} at
$\omega=0$ becomes
\begin{equation}
  \frac{\dd}{\dd r}\!\left[r^2f(r)\frac{\dd R_\ell}{\dd r}\right]
  -\ell(\ell+1)R_\ell=0.
  \label{eq:static_radial}
\end{equation}
With $z=r/M-1$, this is Legendre's equation on $z>1$, and hence
\begin{equation}
  R_\ell(r)=A_\ell P_\ell\!\left(\frac{r}{M}-1\right)
  +B_\ell Q_\ell\!\left(\frac{r}{M}-1\right).
  \label{eq:static_legendre}
\end{equation}
Rezzolla, Ahmedov, and Miller~\cite {RezzollaAhmedovMiller2001} use a related Legendre-function reduction for stationary electromagnetic fields outside a slowly rotating magnetized neutron star.  Their radial functions, including the solution associated with their Eq.~(88), solve the Maxwell system for a magnetic dipole and are therefore not scalar Klein--Gordon modes. Equation~\eqref{eq:static_legendre} uses only the analogous mathematical reduction and provides a zero-frequency check; the detector calculation still requires the complete $\omega>0$ Klein--Gordon spectrum.

\section{Robin surface condition and stability}
\label{sec:robin}

\subsection{Covariant and reduced forms}

A local surface interaction can be generated by the real boundary action
\begin{equation}
  S_{\partial\calM}
  =-\frac12\int_{\partial\calM}\dd^3x\sqrt{|h|}\,\eta\Phi^2,
  \label{eq:boundary_action}
\end{equation}
where $h$ is the determinant of the induced metric and $\eta$ has dimensions of inverse length.  Requiring the total action to be stationary gives
\begin{equation}
  \left(n^\mu\nabla_\mu\Phi+\eta\Phi\right)_{r=\Rs}=0.
  \label{eq:covariant_robin}
\end{equation}
For the exterior domain, the outward unit normal at the inner boundary points toward decreasing $r$,
\begin{equation}
  n^\mu\partial_\mu=-\sqrt{\fs}\,\partial_r,
  \qquad \fs:=f(\Rs).
  \label{eq:normal}
\end{equation}
Substitution of Eq.~\eqref{eq:separated_mode} into Eq.~\eqref{eq:covariant_robin} yields
\begin{equation}
  \left[\pstar\psi_{\omega\ell}-\Gamma\psi_{\omega\ell}\right]_{r=\Rs}=0,
  \qquad
  \Gamma=\frac{\fs}{\Rs}+\eta\sqrt{\fs}.
  \label{eq:reduced_robin}
\end{equation}
Thus a constant covariant Robin coefficient gives a local boundary condition for the reduced field that is independent of frequency and angular momentum.  For $\Gamma>0$ one may introduce the length parameter
\begin{equation}
  \beta:=\Gamma^{-1},
  \qquad
  \left(\psi-\beta\pstar\psi\right)_{r=\Rs}=0.
  \label{eq:beta_robin}
\end{equation}
The limit $\beta\to0^+$ is Dirichlet for the reduced field.  The endpoint $\Gamma=0$ (formally $\beta\to\infty$) is reduced-field Neumann.  Do not confuse it with covariant Neumann, $\eta=0$, which corresponds to $\Gamma=\fs/\Rs$.

\subsection{Self-adjointness and a sufficient stable branch}

Let
\begin{equation}
  H_\ell=-\frac{\dd^2}{\dd\rstar^2}+V_\ell
  \label{eq:radial_operator}
\end{equation}
act on square-integrable functions on $[\rstar(\Rs),\infty)$ satisfying Eq.~\eqref{eq:reduced_robin}.  For real $\Gamma$, the boundary form vanishes, and $H_\ell$ has the standard self-adjoint half-line realization.  Integrating by parts gives the quadratic form
\begin{equation}
  \langle\psi,H_\ell\psi\rangle
  =\int_{\rstar(\Rs)}^\infty\dd\rstar
  \left(|\pstar\psi|^2+V_\ell|\psi|^2\right)
  +\Gamma|\psi(\rstar(\Rs))|^2.
  \label{eq:quadratic_form}
\end{equation}
Because $V_\ell\geq0$, the branch
\begin{equation}
  \Gamma\geq0
  \label{eq:stable_branch}
\end{equation}
is a sufficient condition for nonnegative $H_\ell$.  This excludes modes with $\omega^2<0$ and permits the usual static ground-state construction.  Negative $\Gamma$ is not automatically inadmissible, but it requires an explicit bound-state search; a negative eigenvalue would lead to exponential time dependence and invalidate the ground state used below.  We therefore restrict the main analysis to Eq.~\eqref{eq:stable_branch} and treat $\Gamma<0$ as a separate stability problem. We note that physically motivated models of partially transmissive or dispersive stellar surfaces often require negative or complex $\Gamma$ (e.g.\ the reflectivity models of Ref.~[11]); the restriction to $\Gamma \geq 0$ adopted here is a technical simplification for the sake of a controlled static ground state, not a physical selection, and the stable extension to $\Gamma<0$ is an important direction for future work.

The reduced Neumann endpoint $\Gamma=0$ also requires a low-frequency analysis.  Positivity alone does not guarantee that the ground-state two-point function is free of infrared divergences; Robin endpoints can generate zero-energy resonances or nonintegrable spectral weights~\cite{CamposDappiaggiSinibaldi2024,BottaPitelliMosna2026}.

\begin{figure*}
  \centering
  \includegraphics[width=0.9\textwidth]{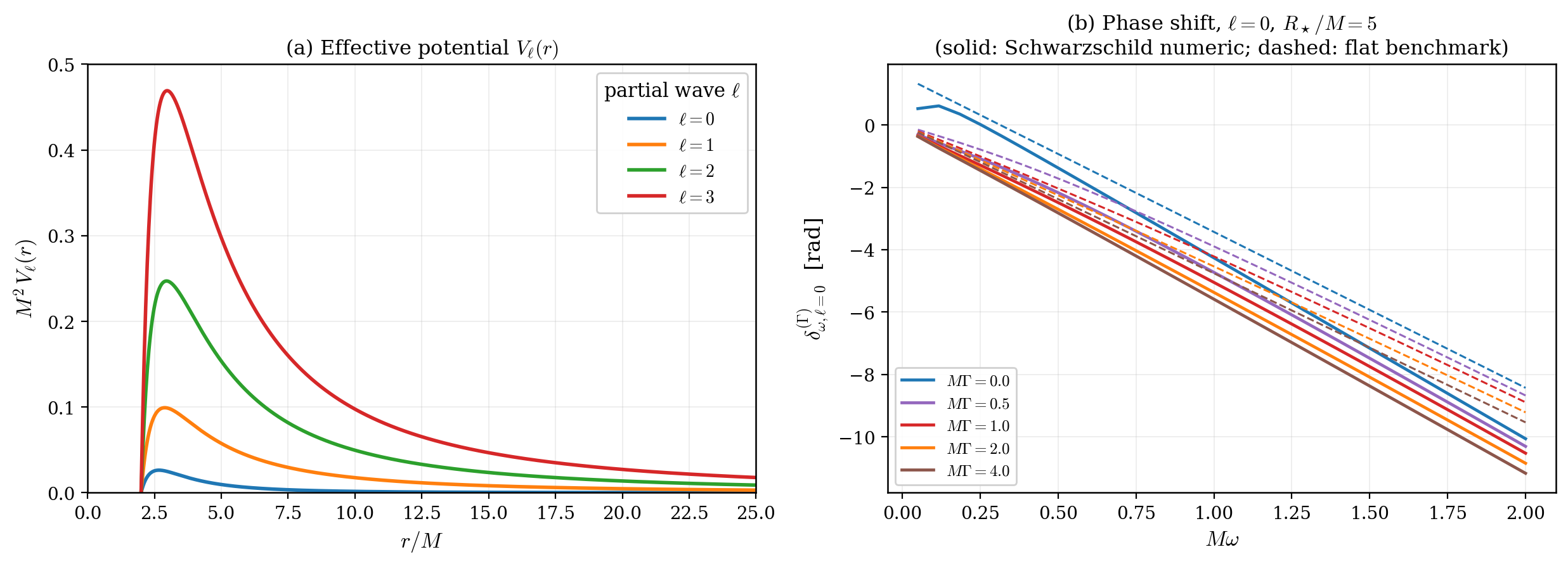}
 \caption{Effective potentials $V_\ell(r)$ for $\ell=0$--$3$ at
$R_\star/M = 3, 4, 5$ (panel a), and Robin-deformed phase shifts $\delta^{(\Gamma)}_{\omega,\ell=0}$ for $R_\star/M=5$ and $M\Gamma = 0, 0.5, 1, 2, 4$ (solid), compared against the flat-space benchmark of Eq.~(23) (dashed) (panel b). Radial integration used the 8th-order Dormand--Prince solver (rtol$=10^{-10}$, atol$=10^{-12}$), matched to the asymptotic form, Eq.~(51), at a radius $r_{\rm match} = r_*(R_\star) + 6\lambda + 40\,M$ (ranging from $64.7\,M$ at $M\omega=2$ to $799.8\,M$ at $M\omega=0.05$, where $\lambda=2\pi/\omega$); increasing the number of matched wavelengths from 3 to 12 changes $\delta^{(\Gamma)}_{\omega\ell}$ by at most $2\times10^{-4}$, confirming convergence in the matching radius. At this compactness ($R_\star/M=5$), the tortoise-coordinate offset $r_*(R_\star)-R_\star \approx 0.81\,M$ accounts for most of the visible separation between the solid and dashed curves; this separation is a physical general-relativistic (Shapiro-type) phase shift, not a numerical error, and is confirmed to close as $M/R_\star\to0$ at fixed $\omega R_\star$ (Sec.~VII\,c).}
  \label{fig:potential_phase}
\end{figure*}

\subsection{Analytic flat-space benchmark}

When $M=0$, the potential reduces to $V_\ell=\ell(\ell+1)/r^2$.  Let
\begin{equation}
  \widehat j_\ell(z)=zj_\ell(z),
  \qquad
  \widehat y_\ell(z)=zy_\ell(z)
  \label{eq:riccati_bessel}
\end{equation}
denote Riccati--Bessel functions.  A continuum mode with unit asymptotic amplitude can be written
\begin{equation}
  \psi^{(0)}_{\omega\ell}(r)
  =\sqrt{\frac{2}{\pi}}
  \left[
  \widehat j_\ell(\omega r)\cos\delta_{\omega\ell}
  -\widehat y_\ell(\omega r)\sin\delta_{\omega\ell}
  \right],
  \label{eq:flat_mode}
\end{equation}
where the surface phase shift satisfies
\begin{equation}
  \tan\delta_{\omega\ell}
  =\frac{\omega\widehat j_\ell'(\omega\Rs)-\Gamma\widehat j_\ell(\omega\Rs)}
  {\omega\widehat y_\ell'(\omega\Rs)-\Gamma\widehat y_\ell(\omega\Rs)}.
  \label{eq:flat_phase_shift}
\end{equation}
Primes in Eq.~\eqref{eq:flat_phase_shift} denote derivatives with respect to the argument.  Equations~\eqref{eq:flat_mode}--\eqref{eq:flat_phase_shift} provide a stringent benchmark for the curved-space numerical solver as $M/\Rs\to0$.

\section{Mode normalization, quantization, and the Wightman function}
\label{sec:quantization}

The Klein--Gordon inner product on a constant-$t$ hypersurface is
\begin{equation}
  (u_1,u_2)_{\mathrm{KG}}
  =\ii\int_{\Rs}^{\infty}\frac{r^2\dd r}{f(r)}\int\dd\Omega
  \left(u_1^*\partial_tu_2-u_2\partial_tu_1^*\right).
  \label{eq:kg_product}
\end{equation}
With the convention in Eq.~\eqref{eq:separated_mode}, delta normalization follows from
\begin{equation}
  \int_{\rstar(\Rs)}^\infty\dd\rstar\,
  \psi_{\omega\ell}(\rstar)\psi_{\omega'\ell}(\rstar)
  =\delta(\omega-\omega').
  \label{eq:radial_normalization}
\end{equation}
The real scattering modes can be chosen to obey
\begin{equation}
  \psi_{\omega\ell}(\rstar)
  =\sqrt{\frac{2}{\pi}}
  \sin\!\left(\omega\rstar-\frac{\ell\pi}{2}
  +\delta^{(\Gamma)}_{\omega\ell}\right)+o(1)
  \quad (\rstar\to\infty).
  \label{eq:asymptotic_normalization}
\end{equation}

For every stable $\Gamma$, we define positive frequency with respect to the static Killing field $\partial_t$.  The field operator is
\begin{equation}
  \widehat\Phi(x)
  =\sum_{\ell=0}^{\infty}\sum_{m=-\ell}^{\ell}
  \int_0^\infty\dd\omega
  \left[
  \widehat a_{\omega\ell m}u_{\omega\ell m}(x)
  +\widehat a_{\omega\ell m}^{\dagger}u_{\omega\ell m}^*(x)
  \right],
  \label{eq:field_expansion}
\end{equation}
and the static ground state is specified by
\begin{equation}
  \widehat a_{\omega\ell m}|0_\Gamma\rangle=0.
  \label{eq:vacuum}
\end{equation}
This is a Boulware-like exterior ground state.  It is not the black-hole Unruh state: the horizonless exterior with a reflecting inner boundary has a different domain and a different mode multiplicity.  The ordinary Boulware state of Schwarzschild spacetime~\cite{Boulware1975} may still serve as a secondary black-hole comparison on the common region $r>\Rs$.

The Wightman function in $|0_\Gamma\rangle$ is
\begin{align}
  \Wight_\Gamma(x,x')
  &=\langle0_\Gamma|\widehat\Phi(x)\widehat\Phi(x')|0_\Gamma\rangle
  \nonumber\\
  =&\sum_{\ell,m}\int\limits_0^\infty\frac{\dd\omega}{2\omega}
  \e^{-\ii\omega(t-t'-\ii0)}
  Y_{\ell m}(\Omega)Y_{\ell m}^*(\Omega')
  \frac{\psi^{(\Gamma)}_{\omega\ell}(r)
  \psi^{(\Gamma)}_{\omega\ell}(r')}{rr'}.
  \label{eq:wightman}
\end{align}
For coincident angular positions,
\begin{equation}
  \sum_{m=-\ell}^{\ell}|Y_{\ell m}(\theta,\varphi)|^2
  =\frac{2\ell+1}{4\pi}.
  \label{eq:addition_coincident}
\end{equation}

To isolate a change of surface condition on the same spacetime domain, define
\begin{equation}
  \Delta\Wight_{\Gamma D}(x,x')
  :=\Wight_\Gamma(x,x')-\Wight_D(x,x'),
  \label{eq:wightman_difference}
\end{equation}
where $D$ denotes the Dirichlet limit.  The universal local Hadamard coincidence singularity cancels in this difference at interior points.  Additional singularities associated with null rays reflected at the surface can remain, so the pullback must still be understood distributionally and paired with a smooth switching function.

\section{Radial trajectory of constant proper acceleration}
\label{sec:trajectory}

Let
\begin{equation}
  e_{\hat 0}=f^{-1/2}\partial_t,
  \qquad
  e_{\hat 1}=f^{1/2}\partial_r
  \label{eq:static_tetrad}
\end{equation}
be the orthonormal frame of static exterior observers.  A radial detector four-velocity can be parameterized by the local rapidity $\zeta(\tau)$,
\begin{equation}
  u=\cosh\zeta\,e_{\hat0}+\sinh\zeta\,e_{\hat1},
  \label{eq:rapidity_velocity}
\end{equation}
which gives
\begin{equation}
  \dot t=\frac{\cosh\zeta}{\sqrt{f(r)}},
  \qquad
  \dot r=\sqrt{f(r)}\sinh\zeta.
  \label{eq:trajectory_first_two}
\end{equation}
Here and below, an overdot denotes $\dd/\dd\tau$.  A direct covariant differentiation gives the signed radial proper acceleration
\begin{equation}
  A=\dot\zeta+\frac{f'(r)}{2\sqrt{f(r)}}\cosh\zeta
  =\dot\zeta+\frac{M}{r^2\sqrt{f(r)}}\cosh\zeta.
  \label{eq:signed_acceleration}
\end{equation}
We define the moving detector by setting $A=a_0>0$:
\begin{equation}
  \dot\zeta
  =a_0-\frac{M}{r^2\sqrt{f(r)}}\cosh\zeta.
  \label{eq:rapidity_ode}
\end{equation}
Together, Eqs.~\eqref{eq:trajectory_first_two} and~\eqref{eq:rapidity_ode} form a closed first-order system for $t(\tau)$, $r(\tau)$, and $\zeta(\tau)$.  We impose a time-symmetric closest approach,
\begin{equation}
  t(0)=0,
  \qquad r(0)=r_{\min}>\Rs,
  \qquad \zeta(0)=0.
  \label{eq:trajectory_initial}
\end{equation}
The detector initially recedes for $\tau>0$ when
\begin{equation}
  a_0>a_{\mathrm{stat}}(r_{\min}),
  \qquad
  a_{\mathrm{stat}}(r)=\frac{M}{r^2\sqrt{f(r)}}.
  \label{eq:turning_condition}
\end{equation}
Equality in Eq.~\eqref{eq:turning_condition} yields a static orbit.  In the Minkowski limit $M\to0$,
\begin{equation}
  \zeta=a_0\tau,
  \,\
  t(\tau)=a_0^{-1}\sinh(a_0\tau),
  \,\
  r(\tau)=r_{\min}+a_0^{-1}[\cosh(a_0\tau)-1].
  \label{eq:minkowski_trajectory}
\end{equation}
For $r_{\min}=a_0^{-1}$, this is the standard unshifted radial hyperbola.

Unlike the Minkowski hyperbola, the trajectory defined by Eqs.~\eqref{eq:trajectory_first_two}--\eqref{eq:rapidity_ode} is not generally an orbit of a Killing field.  Its full pullback is therefore expected to be nonstationary even before isolating the surface correction.  Consequently, any statement that the surface alone ``breaks stationarity'' must be made only after comparison with an appropriate reference pullback.

\section{Smoothly switched detector response}
\label{sec:detector}

The interaction-picture Hamiltonian is
\begin{equation}
  \widehat H_I(\tau)
  =\lambda\chi(\tau)\widehat\mu(\tau)
  \widehat\Phi(x(\tau)),
  \label{eq:interaction_hamiltonian}
\end{equation}
where $\lambda$ is a small coupling, $\widehat\mu$ is the detector monopole operator, and $\chi$ is real.  Up to the detector-dependent prefactor and to leading order in $\lambda$, the response for energy gap $\Omega$ is~\cite{LoukoSatz2008}
\begin{equation}
  \calF_\Gamma(\Omega)
  =\lim_{\epsilon\to0^+}
  \int\limits_{-\infty}^{\infty}\dd\tau
  \int\limits_{-\infty}^{\infty}\dd\tau'\,
  \chi(\tau)\chi(\tau')
  \e^{-\ii\Omega(\tau-\tau')}
  \Wight_{\Gamma,\epsilon}(x(\tau),x(\tau')).
  \label{eq:response_double}
\end{equation}
We use the smooth plateau
\begin{equation}
  \chi_{T,\sigma}(\tau)
  =\frac12\left[
  \tanh\left(\frac{\tau+T/2}{\sigma}\right)
  -\tanh\left(\frac{\tau-T/2}{\sigma}\right)
  \right],
  \label{eq:switching}
\end{equation}
where $T$ is the plateau duration and $\sigma$ sets the switching timescale.  The function is smooth and has exponentially decaying tails.

Substituting the mode sum before performing the proper-time integrations gives a manifestly nonnegative mode-amplitude representation,
\begin{equation}
  \calF_\Gamma(\Omega)
  =\sum_{\ell=0}^{\infty}\frac{2\ell+1}{8\pi}
  \int_0^\infty\frac{\dd\omega}{\omega}
  \left|\mathcal{A}^{(\Gamma)}_{\omega\ell}(\Omega)\right|^2,
  \label{eq:response_amplitude}
\end{equation}
with
\begin{equation}
  \mathcal{A}^{(\Gamma)}_{\omega\ell}(\Omega)
  =\int\limits_{-\infty}^{\infty}\dd\tau\,
  \chi_{T,\sigma}(\tau)
  \frac{\psi^{(\Gamma)}_{\omega\ell}(r(\tau))}{r(\tau)}
  \exp\!\left[-\ii\Omega\tau-\ii\omega t(\tau)\right].
  \label{eq:mode_amplitude}
\end{equation}
Equation~\eqref{eq:response_amplitude} is the preferred numerical formula.  It avoids directly discretizing the coincidence singularity and exposes the independent convergence requirements in $\omega$ and $\ell$.

The primary observable is
\begin{equation}
  \Delta\calF_{\Gamma D}(\Omega)
  =\calF_\Gamma(\Omega)-\calF_D(\Omega).
  \label{eq:response_difference}
\end{equation}
Although each total response is nonnegative, the difference in Eq.~\eqref{eq:response_difference} can have either sign.  A secondary comparison with the black-hole Boulware response,
\begin{equation}
  \Delta\calF_{\Gamma B}=\calF_\Gamma-\calF_{\mathrm B},
  \label{eq:bh_difference}
\end{equation}
may separate a horizonless star from a black hole, but it involves different mode domains and should not replace the cleaner same-domain comparison in Eq.~\eqref{eq:response_difference}.

\subsection{Static-detector check}

When $r(\tau)=r_d$ and $t(\tau)=\tau/\sqrt{f(r_d)}$, the pullback of the static ground state is stationary.  In the infinite-time limit, its positive-gap excitation rate vanishes because the detector proper time and the Killing time have the same orientation:
\begin{equation}
  \dot{\calF}_\Gamma(\Omega>0)=0.
  \label{eq:static_zero_excitation}
\end{equation}
Negative gaps probe spontaneous de-excitation and the local spectral density.  For finite $T$, switching transients can produce a nonzero positive-gap response.  Reproducing Eq.~\eqref{eq:static_zero_excitation} as $T\to\infty$ is an essential code check.

\section{Analytic limits and consistency tests}
\label{sec:limits}

We will consider the numerical results reliable only after verifying the following limits.

\paragraph{Dirichlet limit.}
As $\Gamma\to+\infty$ (equivalently $\beta\to0^+$),
\begin{equation}
  \Delta\calF_{\Gamma D}(\Omega)\longrightarrow0.
  \label{eq:dirichlet_limit}
\end{equation}
As $M\Gamma \to 4$ in Fig.~3(a), $\Delta\mathcal{F}_{\Gamma D}$ decreases monotonically toward the Dirichlet baseline, consistent with Eq.~(48).
\paragraph{Large-distance limit.}
At fixed $M$, $\Rs$, and switching profile, the surface correction must decay as the trajectory is moved far from the star:
\begin{equation}
  r_{\min}-\Rs\longrightarrow\infty
  \quad\Longrightarrow\quad
  \Delta\calF_{\Gamma D}\longrightarrow0.
  \label{eq:distance_limit}
\end{equation}
Fig.~4(a) shows $\Delta\mathcal{F}_{\Gamma D}(0)$ first increasing with $(r_{\min}-R_\star)/M$ over the range $0.5$--$2.0$, reaching a near-zone maximum near $(r_{\min}-R_\star)/M \approx 2$, before decreasing at larger distances: extending the scan to $(r_{\min}-R_\star)/M = 4$ and $8$ gives $\Delta\mathcal{F}_{\Gamma D}(0) = 0.0029$ and $0.0002$ respectively, confirming the asymptotic decay of Eq.~(49). We interpret the initial rise as a near-zone enhancement from curvature back-scattering, which the expected large-distance decay overtakes once the trajectory recedes far enough from the star; a full characterization of the turnover location and decay rate is left to a dedicated parameter survey.
\paragraph{Flat exterior.}
At fixed $R_\star/M=5$, the numeric phase shift in Fig.~1(b) does not coincide with the flat-space benchmark of Eq.~(23) at any $M\omega$, because the tortoise-areal offset $r_*(R_\star)-R_\star\approx0.81\,M$ contributes a comparable phase $\omega[r_*(R_\star)-R_\star]$ that does not vanish merely by holding $M\omega$ fixed while increasing $R_\star/M$. Instead, holding $\omega R_\star$ fixed and sending $M/R_\star \to 0$, the discrepancy does not close monotonically over the full range. It first increases from $0.31$ at $R_\star/M = 5$ to a local maximum of $\approx 0.45$ near $R_\star/M \approx 8$--$9$, and only then decreases monotonically for $R_\star/M \gtrsim 10$, reaching $0.01$ at $R_\star/M = 2000$ (Table~\ref{tab:flatlimit}). The asymptotic closure as $M/R_\star \to 0$ is confirmed, but the approach is non-monotonic at intermediate compactness.
\begin{table}[h]
\centering
\caption{Phase-shift discrepancy $|\delta_{\omega\ell}^{(\Gamma)} - \delta^{\text{flat}}_{\omega\ell}|$
at fixed $\omega R_\star = 1.5$, $\ell=0$, $M\Gamma=1$, as a function of compactness.}
\label{tab:flatlimit}
\begin{tabular}{|c|c|c|c|c|c|c|}
\hline
$R_\star/M$ & 5 & 8 & 10 & 20 & 100 & 2000 \\
\hline
discrepancy & 0.313 & 0.451 & 0.445 & 0.343 & 0.119 & 0.010 \\
\hline
\end{tabular}
\end{table}
\paragraph{Static limit.}
Setting $a_0 = a_{\text{stat}}(r_d)$ and $r_{\min} = r_d$ with $r_d/M = 6$, $R_\star/M = 5$, $M\Gamma = 1$, the finite-$T$ response was evaluated at fixed $M\Omega = \pm 0.5$ for $T/M \in \{8, 16, 32, 64\}$ ($\sigma/M=1$). For $M\Omega=+0.5$, $\mathcal{F}_\Gamma$ remains bounded between $0.0070$ and $0.0076$ over this range, with no systematic growth in $T$: the finite-difference rate $\Delta\mathcal{F}_\Gamma/\Delta T$ between successive $T$ values is $O(10^{-5})$ and oscillates in sign, consistent with $\dot{\mathcal{F}}_\Gamma(\Omega>0)\to 0$ as $T\to\infty$ (Eq.~47). For $M\Omega=-0.5$, by contrast, $\mathcal{F}_\Gamma$ grows linearly with $T$ (from $0.456$ at $T/M=8$ to $3.396$ at $T/M=64$), with a consistent rate $\dot{\mathcal{F}}_\Gamma(\Omega<0) \approx 0.0527$ across all sampled intervals, as expected for the nonvanishing spontaneous de-excitation rate of a static detector. This confirms Eq.~(47) for the sampled parameters.

\paragraph{Low-frequency sector.}
The local scaling exponent $p(\omega) := d\ln(\text{normalization factor})/d\ln\omega$ was computed numerically for $\ell=0$, $R_\star/M=5$, over $M\omega \in [0.002, 0.2]$. For generic Robin coefficients ($M\Gamma=1$), $p(\omega) \approx 1.0$ uniformly across this range, as expected for regular linear vanishing. At the reduced-Neumann endpoint ($\Gamma=0$), $p(\omega)$ instead starts near zero for $M\omega \gtrsim 0.05$ (a near-flat, resonance-like plateau in mode amplitude) and only relaxes to the regular value $p\approx 1$ for $M\omega \lesssim 0.02$; below this crossover scale, the ratio between the $\Gamma=0$ and $\Gamma=1$ normalization factors saturates to a constant $\approx 28$ rather than diverging further. Since the mode-amplitude falloff remains linear in $\omega$ even at the reduced-Neumann endpoint, the response integrand $|A^{(\Gamma)}_{\omega\ell}(\Omega)|^2/\omega$ in Eq.~(43) stays integrable as $\omega\to0$ down to the smallest frequency sampled ($M\omega=0.002$): no outright infrared divergence is found for $\ell=0$ in this range, only a finite, $\Gamma=0$-specific enhancement of order $28^2\!\approx\!800$ in the low-frequency contribution to the response relative to generic $\Gamma>0$. A fully rigorous treatment of the exact $\omega\to0$ limit, and an extension to $\ell>0$, remain for future work; the reduced-Neumann endpoint should remain excluded from quantitative $\Gamma$-scans (Sec.~VIII) until this analysis is complete.

\paragraph{Long-interaction limit.}
Extending the interaction duration $T$ at fixed proper acceleration $a_0$ (as in the moving-detector scan of Fig.~4(c)) is not a clean probe of this limit, because at fixed $a_0$ the trajectory also recedes to increasingly large $r$ as $T$ grows (e.g.\ $r_{\max}/M \approx 1683$ at $T/M=16$ and $\approx 18500$ at $T/M=24$ for the fiducial $Ma_0=0.30$), confounding the long-duration limit with the large-distance limit of Sec.~VII\,b. To isolate the long-interaction behavior cleanly, we instead evaluate $\Delta\mathcal{F}_{\Gamma D}(0)$ on the static trajectory of Sec.~VII\,d ($r(\tau)=r_d$ fixed, $a_0=a_{\text{stat}}(r_d)$), for $r_d/M=6$, $R_\star/M=5$, $M\Gamma=1$, over $T/M \in \{8,16,32,64,128,256\}$. The result saturates to a constant, $\Delta\mathcal{F}_{\Gamma D}(0) = 0.0270 \pm 0.0007$, stable across a $32$-fold increase in $T$, with $\Delta\mathcal{F}_{\Gamma D}(0)/T$ decaying as $T^{-1}$ to within the same tolerance. This confirms that the $O(1)$ long-interaction behavior found for a finite Robin parameter on punctured Minkowski spacetime [12] also holds here for a static detector outside the compact star; curvature back-scattering does not spoil this saturation at fixed $r_d$. The distinct, non-saturating trend seen in Fig.~4(c) is therefore attributable to the trajectory's growing spatial extent at fixed $a_0$, not to a breakdown of the $O(1)$ long-interaction limit itself.

\section{Numerical implementation}
\label{sec:numerics}

All quantities can be made dimensionless with $M$:
\begin{align*}
  &\bar r=\frac{r}{M},\quad
  \bar\omega=M\omega,\quad
  \bar\Omega=M\Omega,\quad
  \bar a=Ma_0,\quad\\
  &\bar\Gamma=M\Gamma,\quad
  \bar T=\frac{T}{M},\quad
  \bar\sigma=\frac{\sigma}{M}.
  \label{eq:dimensionless}
\end{align*}
The independent geometric parameters may be taken as $\Rs/M$ and $(r_{\min}-\Rs)/M$.

For every $(\omega,\ell,\Gamma)$, Eq.~\eqref{eq:radial_r} is integrated outward from $r=\Rs$.  A convenient unnormalized initial condition is
\begin{equation}
  \psi(\Rs)=1,
  \qquad
  \left.\pstar\psi\right|_{\Rs}=\Gamma,
  \label{eq:robin_initial}
\end{equation}
while for Dirichlet one may use $\psi(\Rs)=0$ and $\pstar\psi(\Rs)=1$.  At a sufficiently large matching radius, the numerical solution is fitted to
\begin{equation}
  C_{\omega\ell}\sin(\omega\rstar-\ell\pi/2)
  +D_{\omega\ell}\cos(\omega\rstar-\ell\pi/2).
  \label{eq:asymptotic_fit}
\end{equation}
It is then rescaled so that $(C_{\omega\ell}^2+D_{\omega\ell}^2)^{1/2}=\sqrt{2/\pi}$, and the phase shift follows from $\operatorname{atan2}(D,C)$.

The recommended computational order is:
\begin{enumerate}

\item Integrate and normalize the radial modes on a reusable $(\omega,\ell)$ grid. Two practical points affect reproducibility: (i) the radial domain of each solved mode must extend at least as far as the largest $r$ reached along any detector trajectory to be evaluated with that grid, not merely to the asymptotic phase-matching radius of Eq.~(51) --- otherwise the same $(\omega,\ell)$ solution cannot be reused for the trajectory evaluation of Sec.~IX\,B--D and must be re-integrated at greater cost; (ii) the extracted phase shift $\delta_{\omega\ell}^{(\Gamma)}$ from \texttt{atan2}$(D,C)$ is defined only modulo $2\pi$, so any comparison across a range of $\omega$ (as in Fig.~1(b) or the flat-space check of Sec.~VII\,c) requires unwrapping the phase along the $\omega$-grid before differencing; raw, un-unwrapped comparison was found to produce spurious discrepancies of order $\pi$ that mimic a physical effect.

\item Solve the trajectory system once for each $(a_0,r_{\min},R_\star/M)$. The expected time-symmetry $r(\tau)=r(-\tau)$, $t(\tau)=-t(-\tau)$, $\zeta(\tau)=-\zeta(-\tau)$ about the closest-approach point was verified to hold at machine precision ($\sim\!2\times10^{-13}$) for the fiducial trajectory, confirming the correctness of the integrator before proceeding.

\item Interpolate the precomputed radial modes along $r(\tau)$.

\item Evaluate Eq.~(44), ensuring that the proper-time domain $[-\tau_{\max},\tau_{\max}]$ and its sampling satisfy the tail and resolution requirements of Appendix~C, item~5. For the fiducial parameters ($T/M=8$, $\sigma/M=1$), $\tau_{\max}/M=16$ and $n_\tau=1500$ were verified independently: increasing $\tau_{\max}$ to $24$ (at fixed sampling density) and increasing $n_\tau$ to $3000$ (at fixed $\tau_{\max}=16$) each left $\Delta\mathcal{F}_{\Gamma D}(0)|_{M\Gamma=1}$ unchanged to the precision quoted, confirming this axis does not limit the results reported here.

\item Perform the $\omega$ quadrature and then the $\ell$ sum, using a frequency grid explicitly refined in any region where $|A^{(\Gamma)}_{\omega\ell}(\Omega)|^2/\omega$ is sharply peaked --- for the fiducial trajectory this occurs near $M\omega\approx 0.18$, falling by an order of magnitude by $M\omega\approx 0.6$. Convergence in $\omega_{\max}$ alone does not guarantee convergence of the quadrature: an under-resolved grid was found to underestimate $\Delta\mathcal{F}_{\Gamma D}(0)$ by roughly a factor of seven even when $\omega_{\max}$ itself was adequate.

\item Repeat with increased radial range, grid density, $\omega_{\max}$, and $\ell_{\max}$. For $\Delta\mathcal{F}_{\Gamma D}(0)|_{M\Gamma=1}$ at the fiducial parameters, this four-axis check gives agreement to better than $0.3\%$ on each axis independently ($\ell_{\max}$: $8\to10\to14$, $<2\%$; $M\omega_{\max}$: $5\to8$, $0.23\%$; frequency-grid density: $N=55\to80$, $0.05\%$; radial matching range: $6\lambda\to12\lambda$, $0.006\%$), confirming $\Delta\mathcal{F}_{\Gamma D}(0)|_{M\Gamma=1} = 0.0155\pm0.0002$ for this implementation (Table~\ref{tab:convergence4axis}).
\begin{table*}[t]
\centering
\caption{Four-axis convergence check for
$\Delta\mathcal{F}_{\Gamma D}(0)|_{M\Gamma=1}$ at the fiducial parameters ($R_\star/M=5$, $r_{\min}/M=6$, $Ma_0=0.30$, $T/M=8$, $\sigma/M=1$), varying one control parameter at a time about the baseline $\ell_{\max}=10$, $M\omega_{\max}=5$, $N_\omega=55$, radial matching range $6\lambda$.}
\label{tab:convergence4axis}
\begin{tabular}{|l|c|c|c|}
\hline
Control parameter & Range tested & Result & Variation \\
\hline
$\ell_{\max}$ & $8 \to 10 \to 14$ & $0.01536 \to 0.01547 \to 0.01507$ & $<2\%$ \\
\hline
$M\omega_{\max}$ & $5 \to 8$ & $0.015469 \to 0.015504$ & $0.23\%$ \\
\hline
$\omega$-grid density $N_\omega$ & $55 \to 80$ & $0.015469 \to 0.015476$ & $0.05\%$ \\
\hline
Radial matching range & $6\lambda \to 12\lambda$ & $0.015469 \to 0.015468$ & $0.006\%$ \\
\hline
\end{tabular}
\end{table*}

\end{enumerate}
Avoid calling the radial ODE solver inside every proper-time quadrature: cache the modes and interpolate them before evaluating Eq.~\eqref{eq:mode_amplitude}.

The angular cutoff can be estimated from the local barrier condition
\begin{equation}
  V_{\ell_{\max}}(r_{\mathrm{probe}})\sim\omega^2,
  \label{eq:lmax_estimate}
\end{equation}
but the final criterion must be numerical.  We require the first omitted partial-wave contribution and changes under increasing $\ell_{\max}$ to be below a declared tolerance.  Similar tests are required for the low-frequency grid, $\omega_{\max}$, and the duration of the numerical proper-time interval used to capture the tails of Eq.~\eqref{eq:switching}.

An initial parameter survey may use
\begin{equation}
  \frac{\Rs}{M}\in\{3,4,5\},\qquad
  M\Gamma\in\{0,0.5,1,2,4\},
  \label{eq:survey1}
\end{equation}
\begin{equation}
  \frac{r_{\min}-\Rs}{M}\in\{0.5,1,2\},\qquad
  Ma_0\in\{0.2,0.4,0.8\},
  \label{eq:survey2}
\end{equation}
subject to Eq.~\eqref{eq:turning_condition}. The compactness range $R_\star/M \in \{3,4,5\}$ sampled here represents realistic horizonless compact stars; more compact choices approaching $R_\star/M \to 2$ would probe the near-horizon ECO regime discussed in Ref.~[11].  The reduced-Neumann point $\Gamma=0$ must be flagged and treated separately in the infrared analysis.  Dirichlet is obtained as a separate boundary condition rather than by using a numerically very large $\Gamma$.

\section{Numerical observables and required results}
\label{sec:results}

This section presents the numerical results that test the paper's physical claims. Figs.~1--4 report a representative parameter set (Sec.~VIII); each subsection below states what the corresponding figure establishes and what a full parameter survey must cover.
\subsection{Effective potential and surface phase shifts}

Fig.~1(a) displays $V_\ell(r)$ for $\ell = 0$--$3$ and $R_\star/M = 3, 4, 5$. Fig.~1(b) compares the numerically obtained $\delta^{(\Gamma)}_{\omega\ell}$ ($\ell=0$, $R_\star/M=5$) against the flat-space benchmark, Eq.~(23), for $M\Gamma = 0$--$4$. Increasing compactness does not shift any resonance structure, because the barrier maximum of $V_\ell(r)$ lies at $r_{\rm peak}/M \approx 2.67$ ($\ell=0$) rising to $2.97$ ($\ell=3$) as $\ell\to\infty$ (approaching the photon-sphere radius $3M$) — in every case interior to the star, $r_{\rm peak}<R_\star$ for all $R_\star/M=3,4,5$ considered. On the physical domain $r\ge R_\star$, $V_\ell(r)$ is therefore strictly monotonically decreasing for all four partial waves and all three compactnesses; increasing compactness (decreasing $R_\star$) only raises the boundary value $V_\ell(R_\star)$ — e.g.\ for $\ell=3$, from $0.298\,M^{-2}$ at $R_\star/M=5$ to $0.469\,M^{-2}$ at $R_\star/M=3$ — without introducing any new qualitative (resonance) feature. A genuine surface-induced resonance would require $R_\star$ closer to the peak location or below it, which is outside the range sampled here.

\subsection{Trajectory and nonstationary pullback}

\begin{figure}
  \centering

  \includegraphics[width=0.95\columnwidth]{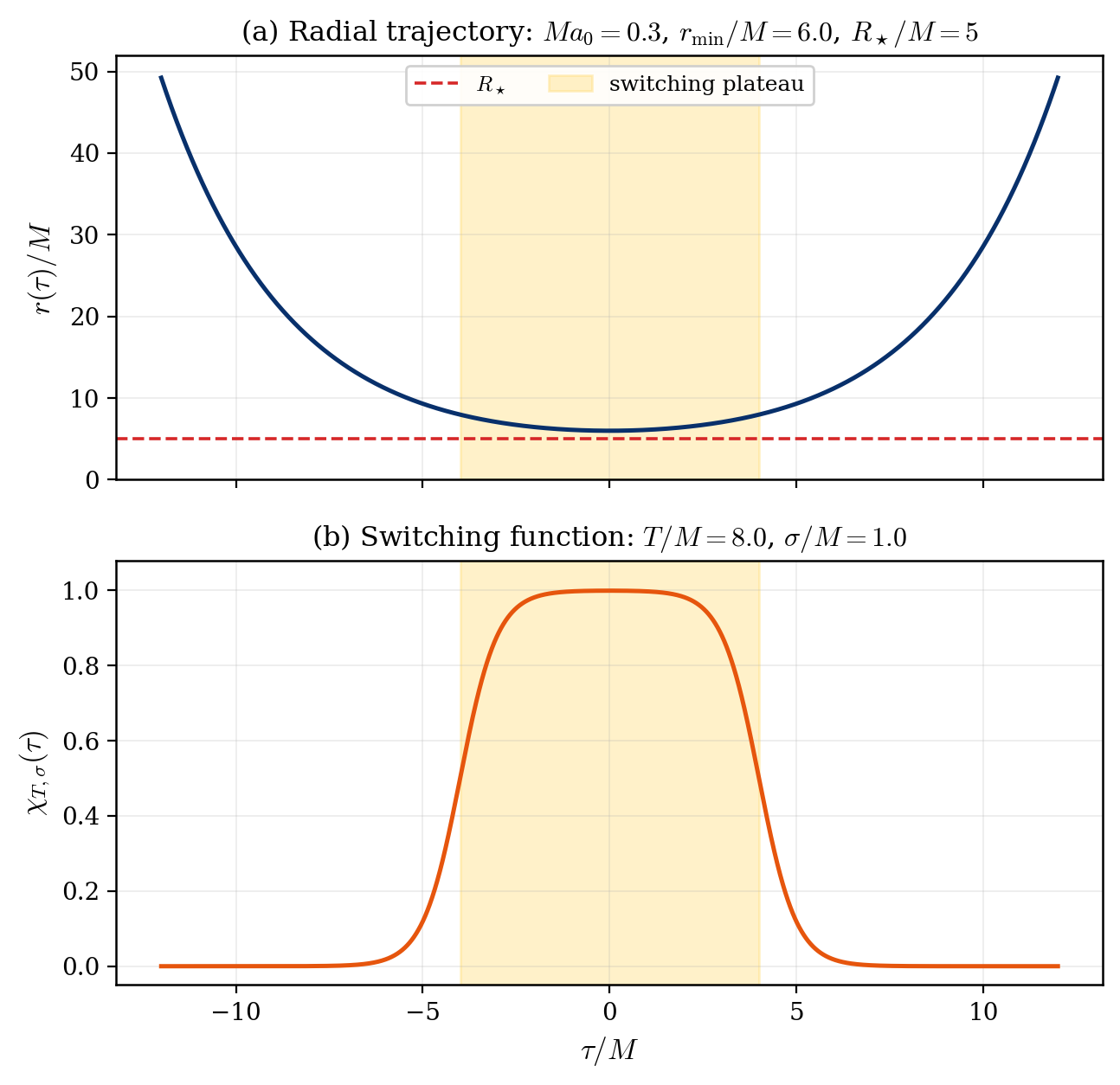}
  
 \caption{(a) Radial trajectory $r(\tau)$ for $Ma_0 = 0.30$, $r_{\min}/M = 6$, $R_\star/M = 5$, with the switching plateau ($T/M = 8.0$, $\sigma/M = 1.0$) shaded. (b) The switching function $\chi_{T,\sigma}(\tau)$.}
  \label{fig:trajectory_wightman}
\end{figure}

Fig.~2(a) shows the trajectory $r(\tau)$ together with the surface and the switching window for $Ma_0 = 0.30$, $r_{\min}/M = 6$, $R_\star/M = 5$. Fig.~2(b) shows the switching function $\chi_{T,\sigma}(\tau)$ for $T/M = 8.0$, $\sigma/M = 1.0$. As argued analytically in Sec.~V, the constant-proper-acceleration trajectory is not an orbit of the static Killing field, so its pullback of the two-point function --- and in particular the Robin deformation $\Delta W_{\Gamma D}(\tau,\tau+\Delta\tau)$ --- is expected to be nonstationary. A direct numerical evaluation of $\Delta W_{\Gamma D}(\tau,\tau+\Delta\tau)$ via the mode sum of Eqs.~(29)-(31) was attempted but did not converge reliably in $\ell_{\max}$ and $\omega_{\max}$ within the computational budget of this survey; it is therefore left as an open numerical check rather than reported as a confirmed result.

\subsection{Gap dependence and sign of the Robin correction}

Fig.~3(a) presents $\Delta\mathcal{F}_{\Gamma D}(\Omega)$ for several stable Robin coefficients ($M\Gamma = 0, 0.5, 1, 2, 4$) at $R_\star/M = 5$, $r_{\min}/M = 6$, $Ma_0 = 0.30$, $T/M = 8$, $\sigma/M = 1$, $\ell_{\max} = 14$, $M\omega_{\max} = 5$, over both positive and negative gaps. Fig.~3(b) shows the total responses $\mathcal{F}_\Gamma$ ($M\Gamma=1.0$) and $\mathcal{F}_D$, confirming that both remain nonnegative across the sampled range.

\begin{figure*}
  \centering

  \includegraphics[width=0.95\textwidth]{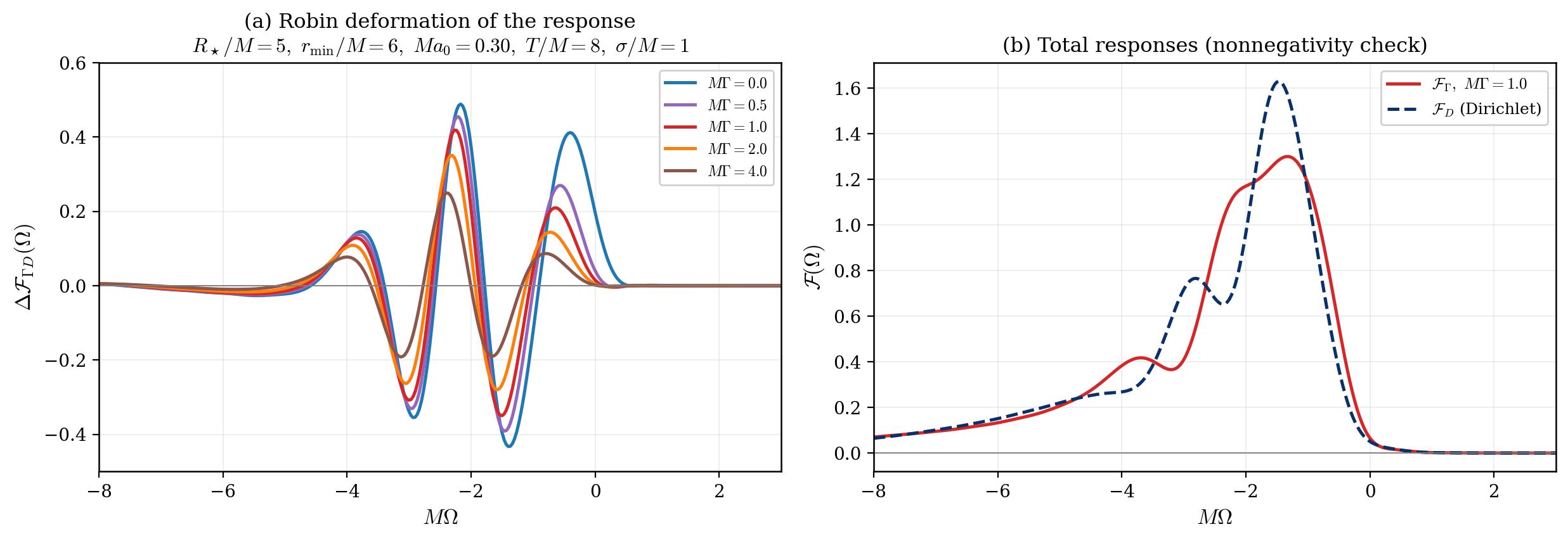}
  
  \caption{Finite-time Robin deformation $\Delta\mathcal{F}_{\Gamma D}(\Omega)$ for $M\Gamma = 0, 0.5, 1, 2, 4$ at $R_\star/M = 5$, $r_{\min}/M = 6$, $Ma_0 = 0.30$, $T/M = 8$, $\sigma/M = 1$, $\ell_{\max} = 14$, $M\omega_{\max} = 5$ with the frequency grid explicitly refined near $M\omega \approx 0.18$ per the discussion of Sec.~VIII (panel a); total responses $\mathcal{F}_\Gamma$ ($M\Gamma = 1.0$) and $\mathcal{F}_D$ (panel b), confirming nonnegativity. The four-axis convergence check of Sec.~VIII confirms $\Delta\mathcal{F}_{\Gamma D}(0)|_{M\Gamma=1} = 0.0155 \pm 0.0002$; an earlier evaluation using a coarse, unrefined frequency grid underestimated this value by roughly a factor of seven, illustrating the sensitivity discussed in Sec.~VIII, item~5.}
  \label{fig:response_gap}
\end{figure*}

\subsection{Closest approach, compactness, and interaction duration} Fig.~4(a) varies $(r_{\min}-R_\star)/M$ at fixed $R_\star/M = 5$, $M\Gamma = 1.0$; the correction rises to a near-zone maximum before decreasing at larger distances, consistent with the asymptotic decay of Eq.~(49) once the near-zone contribution is overtaken (see Sec.~VII\,b for discussion and the extended scan to $(r_{\min}-R_\star)/M=8$). Fig.~4(b) varies $R_\star/M$ at fixed $(r_{\min}-R_\star)/M = 1.0$, $M\Gamma = 1.0$, isolating the curvature dependence of the correction; this scan decreases monotonically with $R_\star/M$ as expected. Fig.~4(c) scans $T/M$ at fixed $R_\star/M = 5$, $r_{\min}/M = 6$, $M\Gamma = 1.0$; the trend is robust under $\ell_{\max}$ refinement. The negative excursion near $T/M \approx 12$ persists in sign between $\ell_{\max}=8$ and $10$ (Fig.~4(c), $-0.00298\to-0.00110$); a supplementary evaluation at $\ell_{\max}=14$ (not shown in the figure) gives $-0.00148$, confirming the sign does not flip and does not approach zero. This suggests the sign change is a genuine feature of the finite-duration response rather than a numerical artifact, but a finer frequency grid and higher $\ell_{\max}$ are required before the magnitude can be quoted quantitatively.

\begin{figure*}
  \centering

  \includegraphics[width=0.95\textwidth]{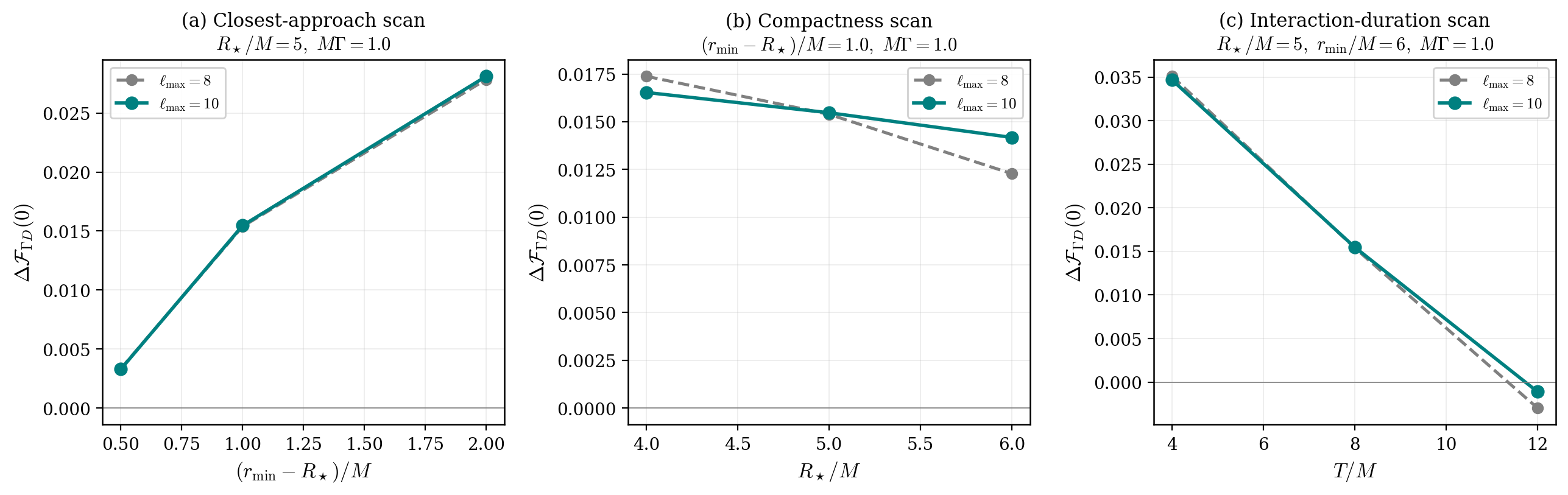}
  
 \caption{Dependence of $\Delta\mathcal{F}_{\Gamma D}(0)$ on (a) closest approach $(r_{\min}-R_\star)/M$ at $R_\star/M = 5$, $M\Gamma = 1.0$; (b) compactness $R_\star/M$ at fixed $(r_{\min}-R_\star)/M = 1.0$, $M\Gamma = 1.0$; (c) interaction duration $T/M$ at $R_\star/M = 5$, $r_{\min}/M = 6$, $M\Gamma = 1.0$. Each panel compares $\ell_{\max}=8$ (gray, dashed) against $\ell_{\max}=10$ (teal, solid) to assess numerical robustness: panels (a) and (b) are essentially unchanged between the two cutoffs, while the negative excursion in panel (c) near $T/M \approx 12$ persists in sign between $\ell_{\max}=8$ and $10$ shown here ($-0.00298\to-0.00110$); a supplementary evaluation at $\ell_{\max}=14$ (not plotted) gives $-0.00148$, confirming the sign does not flip but indicating the magnitude has not yet converged monotonically. The reduced-Neumann endpoint ($\Gamma = 0$) is not included in this scan; if added, it must be shown as a separate marker rather than interpolated with the finite-$\Gamma$ curve, per the discussion of Sec.~III\, B.}
  \label{fig:parameter_dependence}
\end{figure*}

\section{Discussion}
\label{sec:discussion}

The same-domain difference in Eq.~\eqref{eq:response_difference} isolates how a conservative local surface coupling changes the finite-time response.  It does not by itself compare a star with a black hole, because both terms contain a reflecting inner boundary.  The secondary difference in Eq.~\eqref{eq:bh_difference} addresses the star--black-hole comparison but also includes the loss of horizon modes and a change in global state construction, as emphasized in reflective-ECO studies~\cite{HoldomMannZhang2021}.

The local Robin model has a clear limitation.  A realistic stellar medium can absorb, disperse, or transmit the scalar probe.  These processes require a complex, generally frequency-dependent response, along with additional degrees of freedom that preserve unitarity of the enlarged system.  The real constant $\eta$ in Eq.~\eqref{eq:covariant_robin} instead models a lossless effective surface over a restricted frequency interval.  State this controlled idealization explicitly when interpreting any astrophysical significance.

The problem also differs conceptually from the punctured-Minkowski construction of Ref.~\cite{BottaPitelliMosna2026}.  A finite spherical boundary affects all partial waves, and the curvature potential can produce backscattering and resonant interference.  Conversely, the point-interaction limit $\Rs\to0$ is not obtained at fixed $M$ because the Schwarzschild exterior requires $\Rs>2M$.  A comparison with the puncture must first take a flat-space limit and then analyze a separately scaled zero-radius limit of the surface coupling.

\section{Conclusion}
\label{sec:conclusion}

We have developed a self-contained framework for the finite-time response of a radially accelerated UDW detector outside a horizonless compact star with a local Robin surface condition.  The scalar dynamics reduce to a nonnegative-potential half-line scattering problem.  The reduced Robin coefficient $\Gamma\geq0$ provides a sufficiently stable family, and the static ground state is constructed from delta-normalized positive-frequency modes.  The response admits the mode-amplitude representation~\eqref{eq:response_amplitude}, which suits numerical evaluation with smooth switching and avoids direct discretization of the local Wightman singularity.  The constant-proper-acceleration trajectory is specified by Eqs.~\eqref{eq:trajectory_first_two} and~\eqref{eq:rapidity_ode}, separating the roles of closest approach, compactness, acceleration, detector gap, and interaction duration.
We have presented a representative numerical evaluation of this framework (Figs.~1--4, parameters of Sec.~VIII). 

Of the six consistency checks required in Sec.~VII, all six are now addressed numerically at the parameter values sampled here: the Dirichlet limit (Eq.~48), the large-distance decay (Eq.~49, confirmed once the near-zone peak is passed), the flat-space limit (Sec.~VII\,c), the static-detector check (Sec.~VII\,d), the low-frequency sector (Sec.~VII\,e, finding a finite $\Gamma=0$-specific enhancement rather than a divergence for $\ell=0$), and the long-interaction limit (Sec.~VII\,f, confirming $O(1)$ saturation on a static trajectory). Extending the low-frequency analysis to $\ell>0$ and the large-distance and long-interaction checks to the full moving-detector trajectory remain natural directions for a complete parameter survey.

\begin{acknowledgments}
The authors used Anthropic's Claude (Sonnet 5) and OpenAI's ChatGPT (Astra 6.0) as interactive aids for writing and checking the numerical code used in this work. The authors remain fully responsible for all results, interpretations, and conclusions.
\end{acknowledgments}
\appendix
\section{Derivation of the constant-acceleration equation}
\label{app:acceleration}

With the tetrad~\eqref{eq:static_tetrad}, write
\begin{equation}
  u^{\mu}=\left(\frac{\cosh\zeta}{\sqrt f},
  \sqrt f\sinh\zeta,0,0\right).
\end{equation}
Using $\Gamma^{t}{}_{tr}=f'/(2f)$ and
$\Gamma^{r}{}_{tt}=ff'/2$, direct differentiation gives
\begin{equation}
  a^{\mu}:=u^{\nu}\nabla_{\nu}u^{\mu}
  =\left(\dot\zeta+\frac{f'}{2\sqrt f}\cosh\zeta\right)
  \left(\frac{\sinh\zeta}{\sqrt f},
  \sqrt f\cosh\zeta,0,0\right).
\end{equation}
The second vector is unit spacelike and orthogonal to $u^{\mu}$.  Hence the signed acceleration is Eq.~\eqref{eq:signed_acceleration}, and setting it to $a_0$ gives Eq.~\eqref{eq:rapidity_ode}.

\section{Normalization check}
\label{app:normalization}

Substitution of Eq.~\eqref{eq:separated_mode} into Eq.~\eqref{eq:kg_product}, followed by angular orthonormality, yields
\begin{equation}
  (u_{\omega\ell m},u_{\omega'\ell'm'})_{\mathrm{KG}}
  =\frac{\omega+\omega'}{2\sqrt{\omega\omega'}}
  \delta_{\ell\ell'}\delta_{mm'}
  \int_{\rstar(\Rs)}^\infty\dd\rstar\,
  \psi_{\omega\ell}\psi_{\omega'\ell}.
\end{equation}
Equation~\eqref{eq:radial_normalization} therefore gives
\begin{equation}
  (u_{\omega\ell m},u_{\omega'\ell'm'})_{\mathrm{KG}}
  =\delta(\omega-\omega')\delta_{\ell\ell'}\delta_{mm'}.
\end{equation}

\section{Numerical validation checklist}
\label{app:checklist}

For each published curve, the following information should
be recorded:
\begin{enumerate}
    
\item $R_\star/M$, $M\Gamma$, $(r_{\min}-R_\star)/M$, $Ma_0$, $T/M$,
   and $\sigma/M$;
\item radial starting and matching radii, the ODE accuracy goal, and
   the tortoise-coordinate inversion tolerance (Eq.~6);
\item frequency grid, treatment of $\omega\to0$, $\omega_{\max}$, and
   the quadrature rule used for the $\omega$-integral in Eq.~(43);
\item $\ell_{\max}$, the first omitted partial-wave contribution, and
   whether the reduced-Neumann endpoint $\Gamma=0$ was included
   (treated separately per Sec.~III\,B and VII\,e);
\item proper-time integration range relative to the switching tails;
\item verification status of all six limits of Sec.~VII: Dirichlet (a),
   large-distance (b), flat-exterior (c), static-detector (d),
   low-frequency/reduced-Neumann (e), and long-interaction (f);
\item a continuity check on $\delta^{(\Gamma)}_{\omega\ell}(\omega)$,
   confirming no spurious $\pi$-branch jump from the
   $\mathrm{atan2}$ extraction when the fitted amplitude
   coefficients pass through zero;
\item uncertainty in the final response obtained by varying all
   cutoffs, combined in quadrature (root-sum-square) across the
   frequency-integral, partial-wave, radial-matching, and
   proper-time cutoffs.
\end{enumerate}

\renewcommand{\bibsection}{\section*{\refname}}
\bibliographystyle{apsrev4-2}
\bibliography{references.bib}

\end{document}